\pdfoutput=1

\documentclass{article}
 
\def\pprw{8.5in}
\def\pprh{11in}

\usepackage{pbox} 
\usepackage{graphicx}
\usepackage{times}
\usepackage{url}
\usepackage{balance}

\usepackage{paralist} 
\usepackage{booktabs} 
\usepackage{subfigure}
\usepackage{multirow}
\usepackage[pdftex]{hyperref}
\date{}
\begin{document} 
\title{Characterizing the Demographics Behind the \#BlackLivesMatter Movement
\thanks{This is a pre-print of the extended abstract accepted to the AAAI Spring Symposia on Observational Studies through Social Media and Other Human-Generated Content, Stanford, US, March 2016. Please cite the AAAI Spring Symposia version.}}

\author{ 
Alexandra Olteanu \\
EPFL \\
alexandra.olteanu@epfl.ch \\ \\ 
Ingmar Weber \\
QCRI \\
iweber@qf.org.qa \\ \\
Daniel Gatica-Perez \\
Idiap and EPFL	\\
gatica@idiap.ch
}
\maketitle

\begin{abstract} 
The debates on minority issues are often dominated by or held among the concerned minority: gender equality debates have often failed to engage men, while those about race fail to effectively engage the dominant group.  To test this observation, we study the \textbf{\#BlackLivesMatter} movement and hashtag on Twitter---which has emerged and gained traction after a series of events typically involving the death of African-Americans as a result of police brutality---and aim to quantify the \emph{population biases} across user \emph{types} (individuals vs. organizations), and (for individuals) across various demographics factors (\emph{race, gender and age}).  
Our results suggest that more African-Americans engage with the hashtag, and that they are also more active than other demographic groups.  
We also discuss ethical caveats with broader implications for studies on sensitive topics (e.g. discrimination, mental health, or religion) that focus on users. 
\end{abstract}

\section{Introduction}

While the growing number of discussions about minority\footnote{Throughout the paper, by {\em minority} we refer to a group that is subordinate to a more dominant group in society.} issues---including gender~\cite{sandberg-gender-equality}, income~\cite{tale-two-covers}, or race~\cite{anand-giridharadas-ted-inequality}---is good news, empirical evidence suggests that they are held mainly among the discriminated group: women dominate the debate on gender~\cite{female-managers-gender-equality-nhs}, while African-Americans dominate the one on race~\cite{racial-inequality}.  
Although social media has led to a paradigm shift for advocacy by increasing the effectiveness, the speed and the outreach of social campaigns, {\em many still fail to reach far beyond the communities for which they advocate}.

In this paper, we explore this observation in the context of the \#BlackLivesMatter movement~\cite{blacklivesmatter} on Twitter.  We want to gain insights into the level of involvement across user demographics. 
%
What can be said about the demographic composition of the communities engaged in the discussions?  Does the discriminated group dominate the debate? 
Ultimately, engaging diverse stakeholder groups is beneficial for the social campaign's success~\cite{ward2013next}, and knowing the extent to which they contribute to the debate is helpful in learning how to alter the message to appeal to them.   

\smallskip
\noindent \textbf{\#BlackLivesMatter} is a movement (and a hashtag) created after the killing of Trayvon Martin in 2012~\cite{blacklivesmatter}, with over $1,000$ demonstrations being held since then~\cite{blacklivesmatterdem}.  The hashtag has been used during a number of events involving disproportionate police violence against African-Americans, as well as disproportionate reaction of mainstream media in Western Countries when terror attacks happen in these countries compared to when they happen in African countries~\cite{garissa}.

\smallskip
\noindent \textbf{Contributions.} Our main contribution is {\em a demographic characterization of users involved in the \#BlackLivesMatter movement on Twitter.}  
Our findings suggest that African Americans are both more numerous and active than other demographic groups.  
Young females are more likely to actively engage in the debate than men, yet, the proportions of white and African American females are similar.   
Looking at male users, we see a slightly different pattern: young adults still dominate the discussions, but they are largely African Americans.  
Looking at organizations, accounting for about 5\% of profiles, we see a 3 times higher tweeting rate than for individuals. 

To run this study, we also created {\em a collection of about 6,000 Twitter users annotated with demographic information such as race, age, and gender}. In contrast with previous work that reports demographic information by automatically predicting demographic factors for each user based e.g. on their profile picture or name~\cite{minkus2015children,zagheni2014inferring,bakhshi2014faces,mislove2011understanding}, we crowdsourced these annotations.  Although more costly,
we do so to work around known pitfalls of automated user classification such as low recall~\cite{minkus2015children} and classification errors~\cite{yadav2014recognizing}.

\medskip
\noindent \textbf{Limitations and Ethical Challenges.  } 
We note that such an endeavor is not without caveats.  
First, there are intrinsic issues with hashtag-based analyses, and the reliance on a single media platform and public APIs~\cite{tufekci2014big,boyd2012critical}:  
The hashtag we focus on does not cover all the discussions and contributions around the issue at core.  
The movement and hashtag use are recent and we cannot capture the long-term evolution of the demographics behind the core debate. 

Second, there are important ethical challenges~\cite{boyd2012critical}:  Although publicly available, user profile data is inherently sensitive as e.g. users might not anticipate a particular use of their data, especially when created in a context sensitive space and time. This becomes even more delicate when explicitly analyzing their demographic attributes. We discuss these challenges in more depth as we detail our methods and their implications. 
 
\section{Data Collection and Annotation}

\begin{table}
\caption{The basic stats of our dataset.}
\centering\scriptsize\begin{tabular}{l|rr|rr}\toprule
Movement & \multicolumn{1}{c}{Tweets} & \multicolumn{1}{c}{Users} & \multicolumn{1}{|c}{Start Day} & \multicolumn{1}{c}{End Day}  \\ \midrule
\#BlackLivesMatter  & 3.54M & 0.88M & 11.04.2012 &  10.05.2015\\
\bottomrule
\end{tabular}
\label{dataset}
\end{table}

\begin{figure*}[tb]
\includegraphics[width=\textwidth]{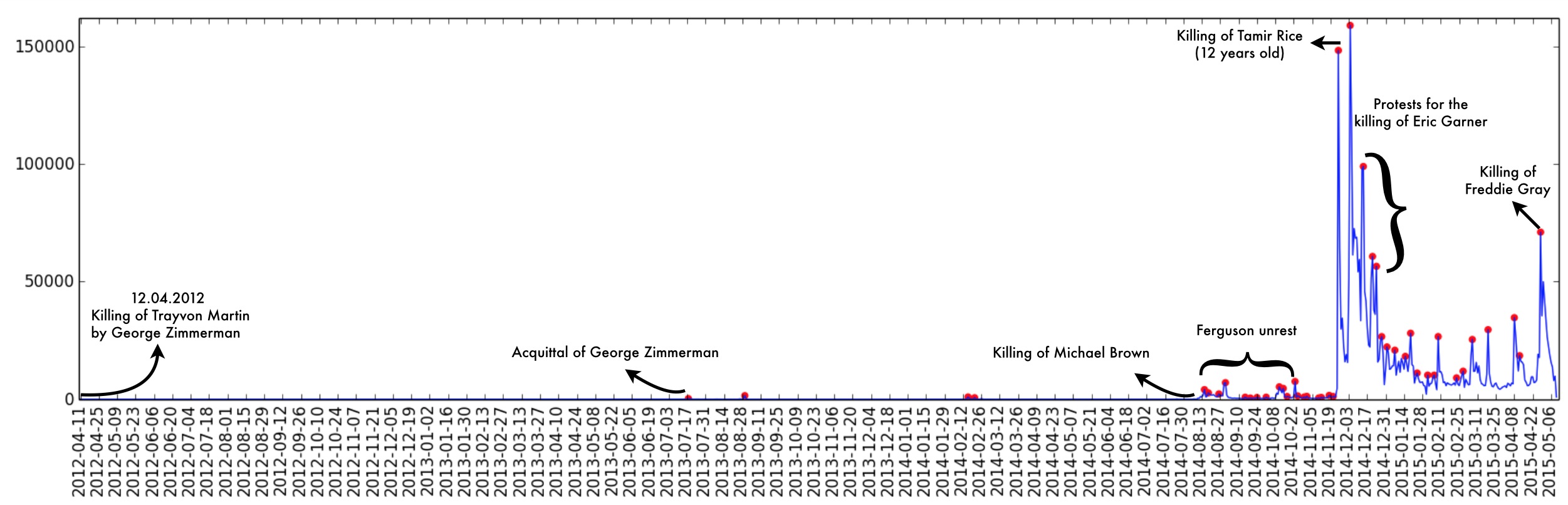}
\centering
\caption{The distribution of the volume of tweets for \#BlackLivesMatter per day over time. }
\label{fig:tweets_dist}
\end{figure*}

\smallskip
\noindent \textbf{The Movement On Twitter.  } The \#BlackLivesMatter hashtag (whose usage over time is highlighted in Figure~\ref{fig:tweets_dist}) was first used on Twitter on April 2012 in relation to the killing of Trayvon Martin~\cite{graeff2014battle}.  
Yet, it grew in a movement only after the acquittal of George Zimmerman (the man who fatally shot Martin) in July 2013,\footnote{http://en.wikipedia.org/wiki/Black\_Lives\_Matter} and got consistent traction after the killing of Michael Brown and with the Ferguson unrest\footnote{http://en.wikipedia.org/wiki/Shooting\_of\_Michael\_Brown}. 
The movement gained momentum after the killing of Tamir Rice,\footnote{http://en.wikipedia.org/wiki/Shooting\_of\_Tamir\_Rice} a 12 year-old school boy, and the decision of a grand jury not to indict the officer that put Eric Garner in a chokehold.\footnote{http://en.wikipedia.org/wiki/Death\_of\_Eric\_Garner} 
Since then, the movement periodically regained public attention with events involving police brutality, including the deaths of Walter Scott\footnote{http://en.wikipedia.org/wiki/Shooting\_of\_Walter\_Scott} and Freddie Gray.\footnote{http://en.wikipedia.org/wiki/Death\_of\_Freddie\_Gray} 

\smallskip
\noindent \textbf{Collecting Tweets. } To gather the tweets published from the day before the first tweet containing the hashtag\footnote{First tweet containing a term obtained via http://ctrlq.org/first/} was posted until 10.05.2015, we crawled Topsy\footnote{http://about.topsy.com/terms-and-conditions/}---the basic figures of the dataset are highlighted in Table~\ref{dataset} and Figure~\ref{fig:tweets_dist}.  To maximize the coverage of our collection, we repeated the crawling with various time window sizes until its' volume converged.  The data was collected in April-May 2015.

\smallskip
\noindent \textbf{Data Collection and Annotation.  } Users data (including public profile data
and crowdsourced annotations) were collected in June 2015.
User profiles were annotated according to the entity behind the Twitter accounts via the crowdsourcing platform Crowdflower\footnote{https://crowdflower.com/}.
We asked the crowd-workers to categorize users as individuals, governmental agencies, NGOs, media, and others; and, then, to categoriza the individuals according to three perceived demographic attributes (race, age and gender).  The crowd-workers were shown automatically generated screenshoots containing the upper part of users public profiles, which included the picture banner, the profile picture, the name and profile description, as well as the last one or two tweets.  The screenshoots were provided via {\em short-lived} URLs in order to limit access to user profile information and minimize the risk of privacy violations. 

We annotated about $6,000$ users from 6 random samples with various characteristics (e.g. from all users, from highly active, from users tweeting about the subject even when the media attention fades away), that are described in the next section.  We showed crowdworkers 5-6 users profiles at a time, out of which one tweet was labeled by one of the authors (gold standard), and used to control the quality of the annotations.  Given that we collect perceived attributes and some of them might be subjective, the user profiles picked to be gold standards were selected to be obvious cases for each of the categories.  For all annotation jobs, we collected at least 3 independent annotations for every user profile and categorization dimension, and kept the majority label. About 100 crowdworkers participated in each task.  Full annotation instructions to be included in the our Data Release.
\section{Exploratory Analysis}
\begin{figure}[t]
\caption{The distribution of number of tweets per user. }
\includegraphics[width=0.75\columnwidth]{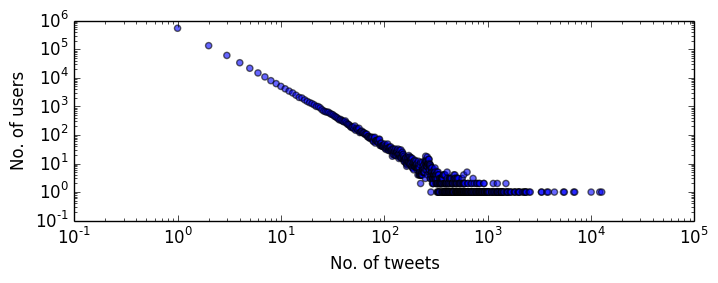}
\centering
\label{fig:user_dist}
\end{figure}

\smallskip
\noindent \textbf{User Distributions.  } The distribution of users according to the number of tweets\footnote{For simplicity, in this paper, by tweets and tweeting we refer to both the act of creating an original tweet, as well as to the act of passing on content, i.e. re-tweeting.} is long tailed (see Figure~\ref{fig:user_dist}), with many users posting only few tweets on the topic (e.g. about 62\% of users have only one tweet in the collection), and only a few users posting in the order of thousands of tweets (only 3 users have more than 10K tweets). This indicates that most of the users participate in the debate only incidentally. For analysis purposes in the rest of the paper, we split users according to their level of activity in three categories: (a) non-active users---$769,231$ users with less than 5 tweets; (b) moderately active users---$96,905$ users with 5 to 25 tweets; and (c) highly active users---$14,033$ users with more than 25 tweets. We make this categorization as we conjecture that the activity w.r.t. a topic is a proxy for a user interest in the topic and her level of involvement, and we want to gain insights into the interplay between activity levels and users demographics.  

Further, we also briefly explore the triggers behind the peaks of attention received by the movement,\footnote{To detect peaks we used a readily available implementation: https://gist.github.com/endolith/250860\#file-peakdet-m} finding that most of the peaks are generated by events involving killing of African-Americans by police in the US (with the debate focusing on the discrimination against African-Americans), see Figure~\ref{fig:tweets_dist}. In addition, the attention peaks for a topic are often indicative of the topic entering and exiting the public debate. When the topic is in the spotlight, a larger community of people tend to get involved in the debate, yet, as the topic fades away, only the concerned community might care. To this end, we define a peak window (or, in other words, the spotlight interval) as a four days interval including the day of the peak, the day before the peak, and two days after the peak. Indeed, using this definition, we found $611,871$ users tweeting in the peak times, as compared to less than half of that number of users being active before the topic ``enters'' or after it ``exits'' the public debate---$268,298$ users. 

\subsubsection{User Characterization}

\begin{table}[tb]
\centering
\caption{Proportion of accounts of organizations vs. accounts of individuals in different samples. Asterisks in the last
row indicate statistically significant differences w.r.t the distribution of all users at $p < 0.01$ (**) and $p < 0.05$ (*)}
\label{tbl:stats-types}
\begin{tabular}{@{}p{0.35in}lllllll}
\toprule
 &   All & Peak & Non   & High    &   Mod.     &  Low      \\
 & Users &  & Peak & Activ. & Activ.  & Activ. \\
 \cmidrule(lr){2-2} \cmidrule(lr){3-4} \cmidrule(lr){5-7}
Org.    & 5.0\% & 4.6\% & 4.9\% & 11.1\% & 5.5\% & 4.2\% \\
Indiv.  & 95.0\% & 95.4\% & 95.1\% & 88.9\% & 94.5\% & 95.8\% \\
\bottomrule
		& 		 &		  &	   		&** 	&*		 &** 
\end{tabular}
\end{table}

To analyze the demographic composition of users involved in the debate we extracted 6 random samples as follows\footnote{Due to technical limitations related to how the screenshots were displayed---resulting in profiles not being shown correctly for annotation---we were able to label only $\sim6000$ users.}: 2,000 users sampled from all users in our dataset; and $5$ other samples of 1,000 users from users tweeting during peak times; users tweeting outside the peak times; highly active users; moderately active users; and from non-active users. The samples were labeled in two rounds: the first annotation task had as main goal the separation of accounts of individuals from accounts of organizations. The second annotation task was designed to categorize accounts of individuals along three demographic criteria: race, gender and age.

\smallskip
\noindent \textbf{Accounts of Organizations.  } 
We first look at the fraction of organization accounts w.r.t those of individuals. 
We notice that the sample drawn from highly active users contains twice as many organization accounts than the other samples. The fraction of organization accounts is higher within more active users (e.g. the sample drawn from moderately active users has a higher fraction of organization accounts than the one drawn from non-active users). This increase comes mostly from a higher fraction of accounts associated with NGOs (7.4\%, 3.6\%, 1\% for highly active, moderately active and non-active users, and 2.2\% across all users) and media organizations, which, however, accounts for the highest fraction among moderately active users (a possible artefact of the fact that media organizations tweet about many topics, while NGOs are typically focused on a handful of causes). Finally, accounts associated with govermental agencies account for less than half a percent in all samples.

\begin{figure}[tb!]
\centering
\subfigure[Distribution of users' age per sample. (best seen in color)]{
\includegraphics[width=0.75\columnwidth]{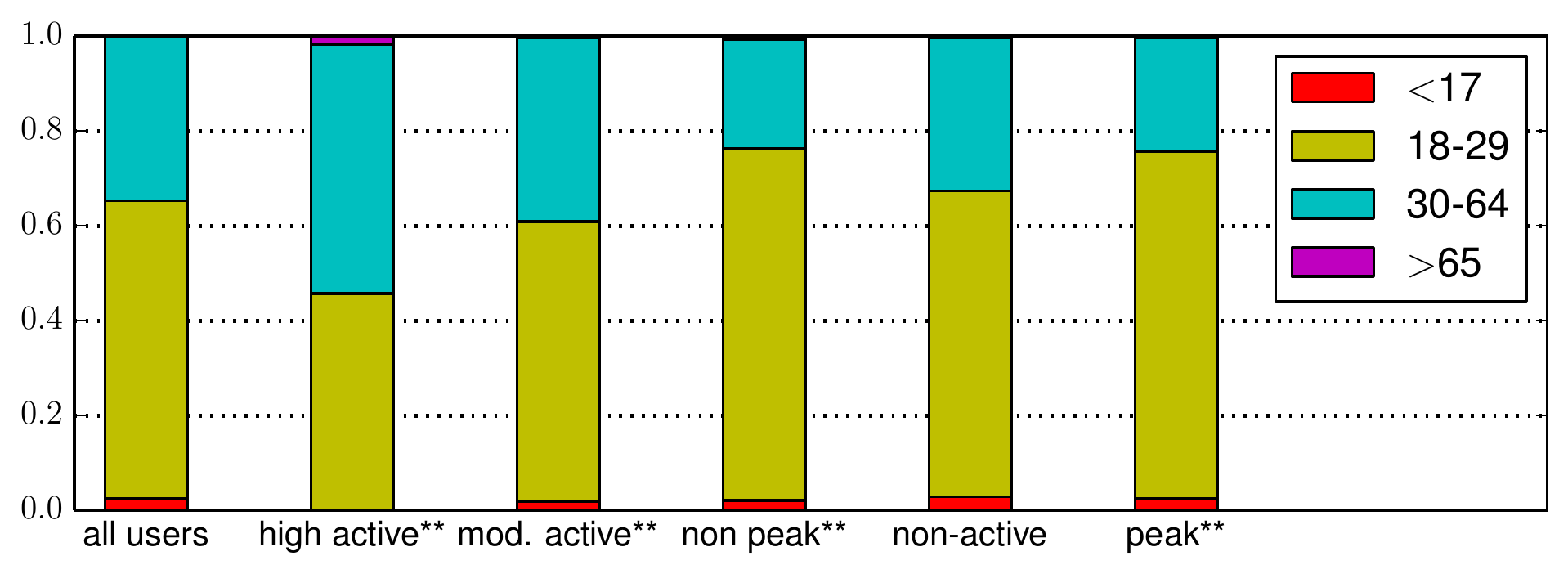}
\centering
}
\subfigure[Distribution of users' race per sample. (best seen in color)]{
\includegraphics[width=0.75\columnwidth]{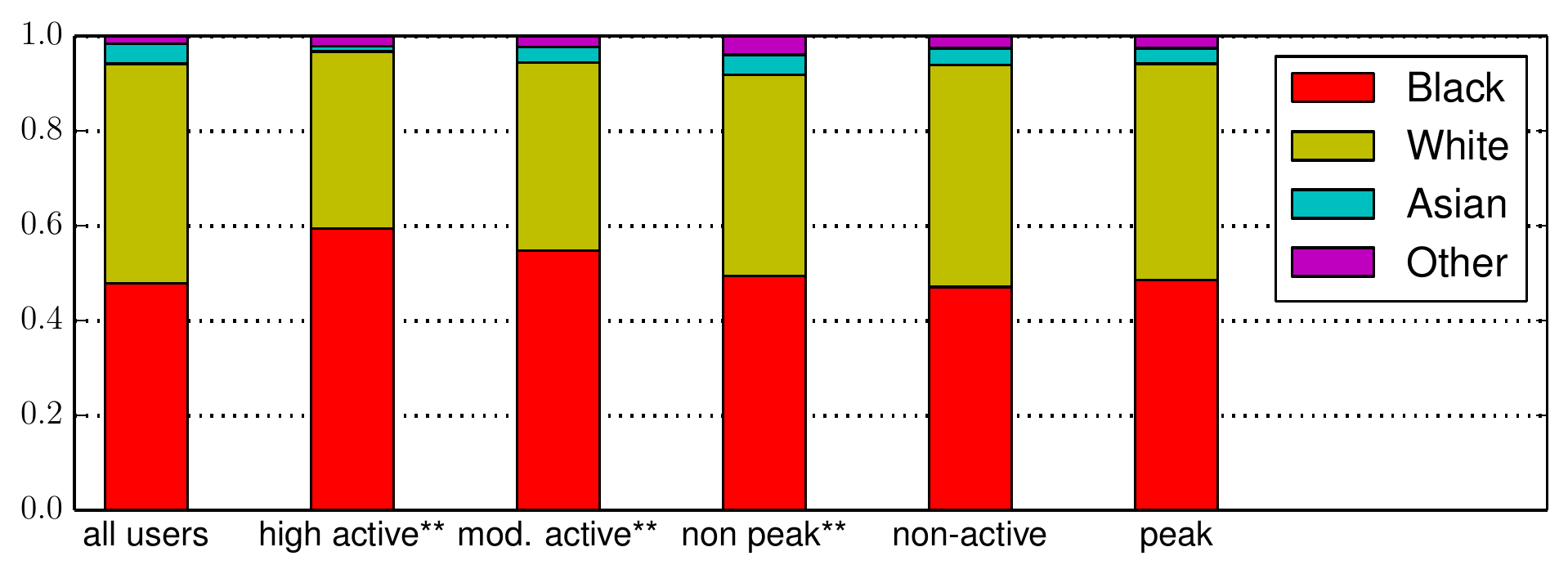}
\centering
}
\subfigure[Distribution of users' gender per sample.]{
\includegraphics[width=0.75\columnwidth]{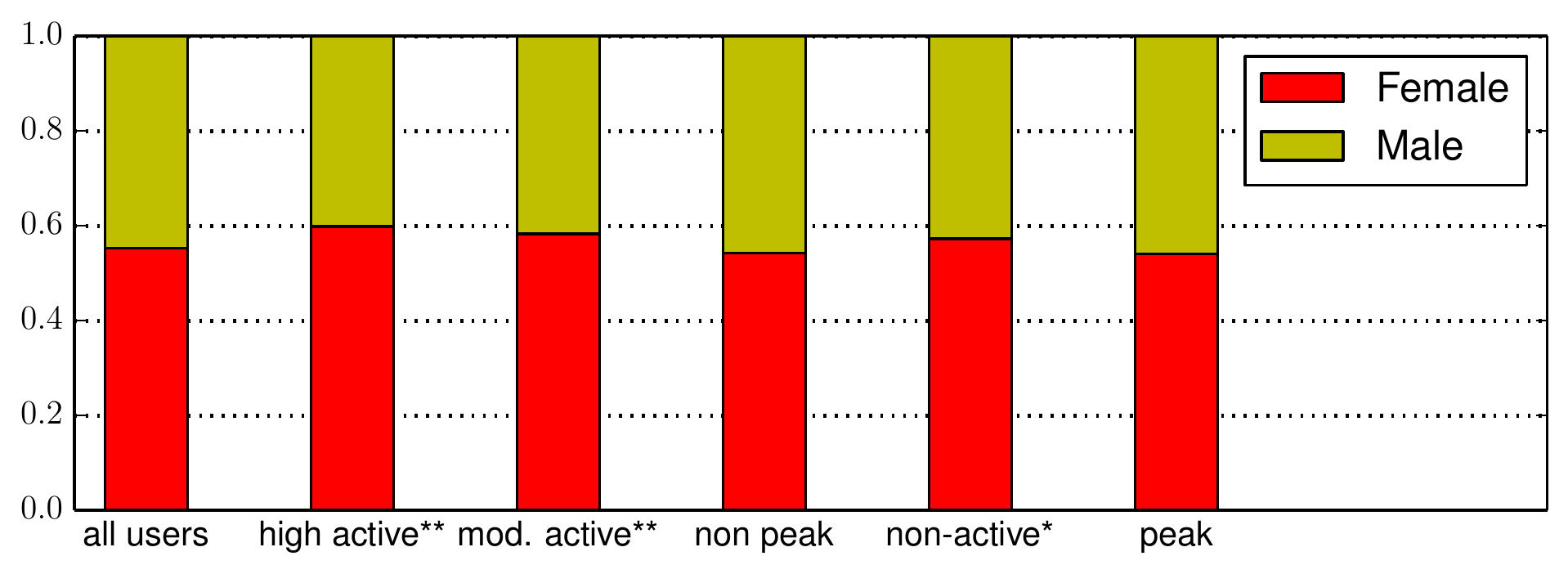}
\centering
}
\caption{Distribution of users by race, age, and gender across samples. Asterisks indicate statistically significant differences w.r.t. the distribution of all users at $p < 0.01$ (**) and $p < 0.05$ (*).}
\label{fig:dist-dem} 
\end{figure}

\smallskip
\noindent \textbf{User Demographics.  } ({\em Age}) For accounts of individuals, we look at the distribution of user demographic factors. In Figure~\ref{fig:dist-dem}(a) we see that the fraction of young adults is lower in the sample of highly active users, while the fraction of adults between 30 to 64 years old is lowest outside the peak times. This could indicate that users in the later category are more active during peak times when the topic is in the public spotlight. ({\em Race}) Figure~\ref{fig:dist-dem}(b) shows how the user distribution across the racial groups varies by sample. We notice that the fraction of African-Americans is the highest within the sample of highly active users, being the smallest among the non-active users or during peak times. ({\em Gender}) Finally, in Figure~\ref{fig:dist-dem}(c) we see that the users distribution according to their gender is stable across samples.  

\begin{figure}[tb!]
\centering
\subfigure[Distribution of male users as a function of age and race. All cells sum to 100\%.]{
\includegraphics[width=0.9\columnwidth]{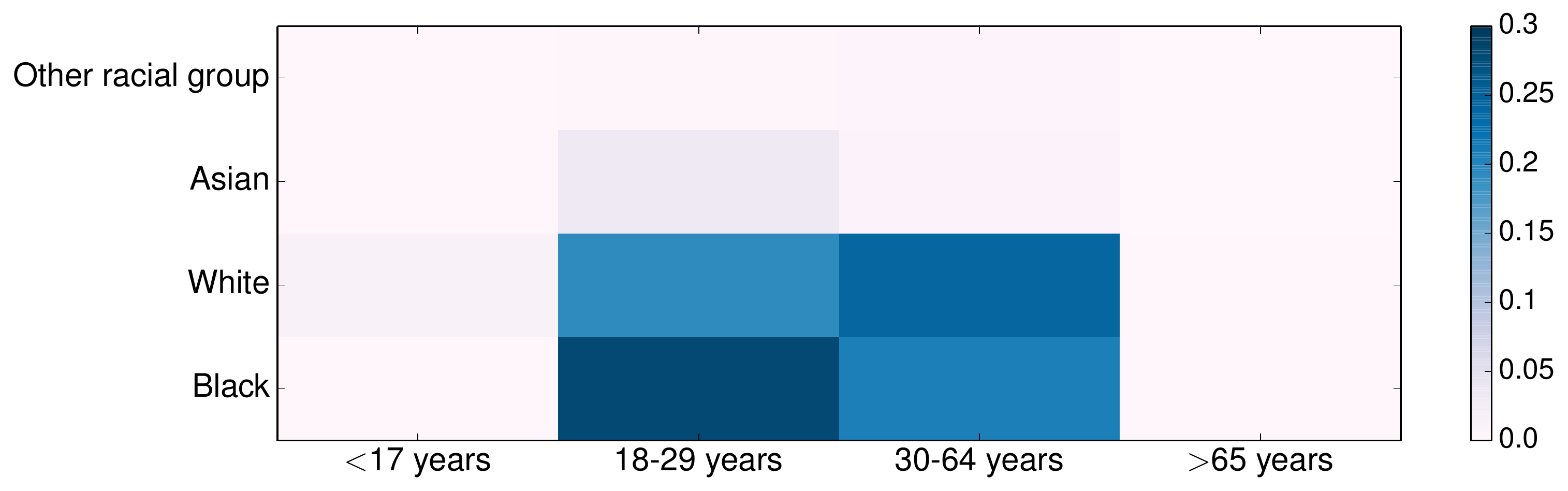}
\centering
}
\subfigure[Male to female ratio. Red indicates a higher fraction of female users than in the overall distribution ($\sim0.78$ marked by white in the colorbar), while blue indicates a higher fraction of men. The percentages represent the distribution of all users. (best seen in color)]{
\includegraphics[width=0.9\columnwidth]{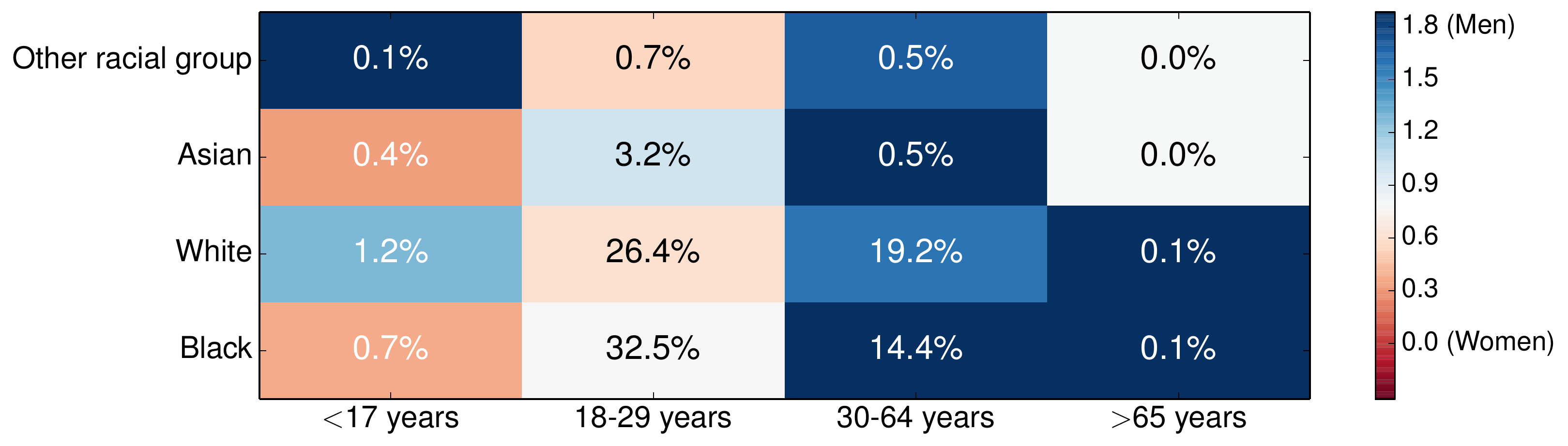}
\centering
}
\caption{Distribution of race and age for female vs. male users.}
\label{fig:dist-heatmap}
\end{figure}

Next, we looked at the distribution of users across age and race for each gender (we omit the figure for women as it follows similar patterns)---see Figure~\ref{fig:dist-heatmap}\footnote{We compute these stats based on those users annotated along all three demographic factors as in some cases only one or two of these factors might be perceptible based on profile information.}.  
We notice that the most active users are white and African-American adults between 18 to 64 years old.  However, while for male African-American users the fraction of young adults (18 to 29 years old) is higher, for white users it is lower.  Looking at the differences between women and men (Figure~\ref{fig:dist-heatmap}(b)), we see that women younger than 29 years old are more active than men in the same age category, while for users older than 30 years old, men tend to tweet more about the movement.

\smallskip
\noindent \textbf{User Involvement.   } Next we wanted to see if users belonging to specific demographic groups are more vocal than others, or, in other words, if they produce more content on average. First, we find that organizations are more active than individuals (7:2). Then, depending on the demographic criteria, we see that: (a) African-Americans are most active, followed closely by white users; (b) women are more active than men (3.8:2.6); and (c) adults between 30 and 64 years old are the most active, followed by young adults (3.9:2.6:2).

\section{Concluding Remarks}

We started the study after one of the related events---the shooting of Walter Scott---and based on empirical evidence we hypothesized that the debate would be hold mainly among African-Americans.  While our findings support this premise, with African-Americans being the largest group (going up to 60\% among highly active users), overall, whites make up about 40\% of individuals and Asians 4\%. Future work naturally includes an analysis of demographic factors across various movements related to minority groups issues in order to validate and broaden the observations we make here. 

\noindent \textbf{Parting Thoughts on Ethics. } Although important, studies looking at various online media to understand the public opinion and the different narratives on minority groups issues across stakeholders are scant, but growing~\cite{chou2015race}.  One reason, the limits in collecting and annotating users accurately and at scale (either manually or automatically).  Yet, as we learn to work around these limits, {\em we also ought to develop protocols to mindfully study such user collections while protecting the users}.  

\smallskip

\noindent \textbf{Data Release. } The tweets collection, including tweet ids and peaks will be made available for research purposes at {\em http://crisislex.org/}. The list of annotated users along with profile information will only be shared upon signing a commitment for \textbf{not} using this data to study users in isolation or to single them out for their demographic attributes or opinions.

\noindent \textbf{Acknowledgements. } We are grateful to Carlos Castillo for his feedback on an early draft of this project.   

\bibliographystyle{acm}
\bibliography{paper}

\begin{thebibliography}{10}

\bibitem{blacklivesmatter}
Black lives matter. [14-april-2015].
\newblock http://blacklivesmatter.com/contact/.

\bibitem{tale-two-covers}
Black lives matter; a tale of two covers. [14-april-2015].
\newblock
  http://www.nbcnews.com/news/nbcblk/black-lives-matter-tale-two-covers-n339796.

\bibitem{blacklivesmatterdem}
Black lives matter demonstrations. [20-sept-2015].
\newblock https://www.elephrame.com/textbook/protests.

\bibitem{racial-inequality}
Can we talk about race? a few rules of engagement. [14-april-2015].
\newblock
  http://articles.baltimoresun.com/2006-08-01/news/0608010135\_1\_racial-inequality-political-change-problem-of-racial.

\bibitem{sandberg-gender-equality}
Gender equality won't happen unless men speak up. [14-april-2015].
\newblock http://edition.cnn.com/2013/04/17/business/sandberg-gender-equality/.

\bibitem{garissa}
Paying attention to {G}arissa. [14-april-2015].
\newblock
  http://www.ethanzuckerman.com/blog/2015/04/04/paying-attention-to-garissa/.

\bibitem{anand-giridharadas-ted-inequality}
Seven signs you are clueless about income inequality. [14-april-2015].
\newblock http://fortune.com/2015/03/20/anand-giridharadas-ted-inequality/.

\bibitem{female-managers-gender-equality-nhs}
What's missing from the debate about women leaders in the nhs? men.
  [14-april-2015].
\newblock
  http://www.theguardian.com/healthcare-network/2014/jan/08/female-managers-gender-equality-nhs.

\bibitem{bakhshi2014faces}
{\sc Bakhshi, S., Shamma, D.~A., and Gilbert, E.}
\newblock Faces engage us: Photos with faces attract more likes and comments on
  instagram.
\newblock In {\em CHI\/} (2014).

\bibitem{boyd2012critical}
{\sc Boyd, D., and Crawford, K.}
\newblock Critical questions for big data: Provocations for a cultural,
  technological, and scholarly phenomenon.
\newblock {\em Information, communication and society\/} (2012).

\bibitem{chou2015race}
{\sc Chou, S.}
\newblock Race and the machine: Re-examining race and ethnicity in data mining.
  2015.
\newblock http://www.sophiechou.com/papers/chou\_racepaper.pdf.

\bibitem{graeff2014battle}
{\sc Graeff, E., Stempeck, M., and Zuckerman, E.}
\newblock The battle for `{T}rayvon {Ma}rtin': Mapping a media controversy
  online and off-line.
\newblock {\em First Monday\/} (2014).

\bibitem{minkus2015children}
{\sc Minkus, T., Liu, K., and Ross, K.~W.}
\newblock Children seen but not heard: When parents compromise children's
  online privacy.
\newblock In {\em WWW\/} (2015).

\bibitem{mislove2011understanding}
{\sc Mislove, A., Lehmann, S., Ahn, Y.-Y., Onnela, J.-P., and Rosenquist,
  J.~N.}
\newblock Understanding the demographics of twitter users.
\newblock {\em ICWSM\/} (2011).

\bibitem{tufekci2014big}
{\sc Tufekci, Z.}
\newblock Big questions for social media big data: Representativeness, validity
  and other methodological pitfalls.
\newblock In {\em ICWSM\/} (2014).

\bibitem{ward2013next}
{\sc Ward, J.~A.}
\newblock The next dimension in public relations campaigns: A case study of the
  it gets better project.
\newblock {\em Public Relations Journal\/} (2013).

\bibitem{yadav2014recognizing}
{\sc Yadav, D., Singh, R., Vatsa, M., and Noore, A.}
\newblock Recognizing age-separated face images: Humans and machines.
\newblock {\em PloS one\/} (2014).

\bibitem{zagheni2014inferring}
{\sc Zagheni, E., Garimella, V. R.~K., Weber, I., et~al.}
\newblock Inferring international and internal migration patterns from twitter
  data.
\newblock In {\em WWW companion publication\/} (2014).

\end{thebibliography}

\end{document}